\documentclass[twocolumn,showpacs,preprintnumbers,prd,superscriptaddress,nofootinbib]{revtex4-1}

\usepackage{amsmath}
\usepackage{amsfonts}
\usepackage{amssymb}
\usepackage{graphicx}
\usepackage{color}

\usepackage{hyperref}
\usepackage{booktabs}
\usepackage{hyperref}
\usepackage{cleveref}
\usepackage{color}
\Crefrangeformat{equation}{Eqs. (#3#1#4)--(#5#2#6)}
\Crefname{equation}{Eq.}{Eqs.}
\Crefname{figure}{Fig.}{Figs.}
\Crefname{section}{Sec.}{Secs.}

\begin{document}

\title{Rational approximations of $f(R)$ cosmography  through Pad\'e polynomials}

\author{Salvatore Capozziello}
\email{capozziello@na.infn.it}
\affiliation{Dipartimento di Fisica, Universit\`a di Napoli  ``Federico II'', Via Cinthia, I-80126, Napoli, Italy.}
\affiliation{Istituto Nazionale di Fisica Nucleare (INFN), Sez. di Napoli, Via Cinthia 9, I-80126 Napoli, Italy.}
\affiliation{Gran Sasso Science Institute, Via F. Crispi 7, I-67100, L' Aquila, Italy.}

\author{Rocco D'Agostino} \email{rocco.dagostino@roma2.infn.it}
\affiliation{Dipartimento di Fisica, Universit\`a degli Studi di Roma ``Tor Vergata'', Via della Ricerca Scientifica 1, I-00133, Roma, Italy.}
\affiliation{Istituto Nazionale di Fisica Nucleare (INFN), Sez. di Roma ``Tor Vergata'', Via della Ricerca Scientifica 1, I-00133, Roma, Italy.}

\author{Orlando Luongo}	\email{luongo@na.infn.it}
\affiliation{Dipartimento di Fisica, Universit\`a di Napoli  ``Federico II'', Via Cinthia, I-80126, Napoli, Italy.}
\affiliation{Istituto Nazionale di Fisica Nucleare (INFN), Sez. di Napoli, Via Cinthia 9, I-80126 Napoli, Italy.}
\affiliation{School of Science and Technology, University of Camerino, I-62032, Camerino, Italy.}
\affiliation{Department of Mathematics and Applied Mathematics, University of Cape Town, Rondebosch 7701,
Cape Town, South Africa.}
\affiliation{Astrophysics, Cosmology and Gravity Centre (ACGC), University of Cape Town, Rondebosch 7701,
Cape Town, South Africa.}

\begin{abstract}
We consider high-redshift $f(R)$ cosmography adopting the technique of polynomial reconstruction. In lieu of considering Taylor treatments, which turn out to be non-predictive as soon as $z>1$, we take into account the Pad\'e rational approximations which consist in performing expansions converging at high redshift domains. Particularly, our strategy is to reconstruct $f(z)$ functions first, assuming the Ricci scalar to be invertible with respect to the redshift $z$. Having the so-obtained $f(z)$ functions, we invert them and we easily obtain the corresponding $f(R)$ terms. We minimize error propagation, assuming no errors upon redshift data. The treatment we follow naturally leads to evaluating curvature pressure, density and equation of state, characterizing the universe evolution at redshift much higher than standard cosmographic approaches. We therefore match these outcomes with small redshift constraints got by framing the $f(R)$ cosmology through Taylor series around $z\simeq 0$. This gives rise to a calibration procedure with small redshift that enables the definitions of polynomial approximations up to $z\simeq 10$. Last but not least, we show discrepancies with the standard cosmological model which go towards an extension of the $\Lambda$CDM paradigm, indicating an effective dark energy term evolving in time.  We finally describe the evolution of our effective dark energy term by means of basic techniques of data mining.   
\end{abstract}

\maketitle

\section{Introduction}

According to the  current observations, the universe started accelerating at a given time and it is today dominated by an exotic accelerating component, called  dark energy \cite{Copeland06,Bamba12,Davis16,Joyce16,Kleidis16}. Attempts to determine either dark energy nature or its micro-physics have been so far unsuccessful. Essentially, the standard cosmological puzzle suggests a constant dark energy modeled through a cosmological constant  $\Lambda$  \cite{Carroll92,SNe,Sahni00,WMAP9,Planck15}. The $\Lambda$ origin is supposed to come from quantum fluctuations at the very beginning of the universe's evolution. Although attractive and straightforward, the corresponding model, dubbed the $\Lambda$CDM paradigm, does not predict cosmological constant's magnitude in agreement with quantum field theory. Furthermore, the magnitude of $\Lambda$ is even surprisingly close to the matter value today, indicating a unexpected coincidence between matter and dark energy \cite{Weinberg89,Zlatev99,Sahni02}. These problems, together with other astronomical indications, suggest that dark energy may slightly evolve in terms of the cosmic redshift $z$, leading to a negative pressure which becomes dominant over standard gravity after a precise time.

Instead of modifying the net equation of state (EoS) with \emph{ad hoc} assumptions on the form of dark energy, plausible modifications of Einstein's gravity have been proposed as energy scales increase. Under this hypothesis, one can postulate modified gravitational theories aiming to describe the dark energy effects by means of \emph{first principles}. Among all, the Hilbert-Einstein action can be generalized by  replacing the Ricci scalar, $R$, with generic $f(R)$ functions \cite{Sotiriou10,De Felice10, Odintsov,Capozziello11}. The great disadvantage of this approach is that the function $f(R)$ is not known \emph{a priori}. Approaches towards the determinations of $f(R)$ have been severely discussed \cite{Hu07,Starobinsky07,Appleby07,Tsujikawa08,Cognola08}, with particular regards to matching cosmic data with the reconstructed $f(R)$. Unfortunately, expanding $f(R)$ in Taylor series as $R$ tends to its observable late time value does not seem enough to describe either dark energy or dark matter at different scales. This caveat is a consequence of the expansion itself which is performed in a short range of redshifts only.

In this paper, we thus propose the use of rational approximations to reconstruct the form of $f(R)$ using constraints coming from background cosmology. We motivate such a scheme because rational approximants are in general capable of expanding the range of redshifts which are not covered by Taylor series. Hence, physically speaking the use of rational approximations overcomes bad convergence issues at high redshift domains. We here employ the Pad\'e rational approximants which are reaching great interest during the last times \cite{Gruber13}. The Pad\'e approximations represent a treatment which is here developed to get a class of $f(R)$ models compatible with cosmic data at higher redshifts.

To figure this out, we consider the function $f(z)$ which corresponds to the function $f(R)$, with the recipe $R=R(z)$. We thus frame the numerical evolution of modified Friedmann equations and we get the numerical behaviour of $f(z)$. Since $R$ is a function of the Hubble rate, we expand it either in Pad\'e series or in Taylor expansions and we even compare the outcomes coming from the two different approaches. As Pad\'e rational orders\footnote{As it will be clarified later, Pad\'e series employs two orders instead than one, as Taylor treatments do.} we consider the ones which turn out to be more compatible with cosmic data. In particular, to select the orders, we consider the ones which reduce error bars in numerical analyses which make use of cosmological surveys. To guarantee that the Pad\'e rational approximations have been well-constructed we take cosmological constraints coming from cosmography. In such a way, we fix bounds over the numerical $f(z)$ which are compatible with state-of-the-art cosmology at late times. This procedure represents a calibration of our high-redshift approximations with at lower redshifts. Once reconstructed the form of $f(z)$, we turn back to $f(R)$ and we define the corresponding cosmology associated to the modified Friedmann equations. Moreover, we find through simple tools of statistical data mining how to better formalize the effective dark energy evolution at redshifts $z\geq 1$. To this end, we reproduce the effective dark energy and its properties, showing slight departures with respect to the standard $\Lambda$CDM model. 

The structure of the paper is the following: in \Cref{sec:f(R) theories}, we introduce the $f(R)$ theories of gravity and the modified Friedmann equation based on the assumption of the cosmological principle. In \Cref{sec:Pade}, we review the method of the Pad\'e polynomials upon which we will build our analysis. In \Cref{sec:cosmography}, we show how to reconstruct the form of $f(R)$ in a model-independent way by means of the cosmographic method. In \Cref{sec:cosmology}, we study the cosmological properties of the obtained $f(R)$ model. In \Cref{sec:Taylor}, we compare the differences between the standard Taylor approach and the Pad\'e approximation. We also describe the evolution of our effective dark energy term with its implications in the redshift domain that we considered. 
Finally, in \Cref{sec:conclusions} we draw conclusions and perspectives of our formalism.


\section{$f(R)$ gravity}
\label{sec:f(R) theories}

The standard gravitational Lagrangian makes use of the first order invariant Ricci scalar. Replacing the Lagrangian with a generic function of $R$, i.e. $f(R)$, implies that the gravitational action leads to a new class of models, by means of: \cite{Capozziello,altri}:

\begin{equation}
S=\int d^4x \sqrt{-g}\left[f(R)+\mathcal{L}_m\right]\,,
\label{eq:action f(R)}
\end{equation}

\noindent where $g$ is the metric determinant, whereas $\mathcal{L}_m$ represents the matter Lagrangian\footnote{Here, we use units such that $8\pi G=1=c$.}.  Varying the action with respect to the metric, we obtain the field equations:
\begin{equation}
R_{\mu\nu}-\dfrac{1}{2}g_{\mu\nu}R=T_{\mu\nu}^{(curv)}+T_{\mu\nu}^{(m)}\,,
\label{eq:field eqs}
\end{equation}
in which $T_{\mu\nu}^{(m)}$ is the energy-momentum tensor of matter and
\begin{equation}
T_{\mu\nu}^{(curv)}=\dfrac{1}{f'}\bigg[\dfrac{1}{2}g_{\mu\nu}(f-Rf')+(g_{\alpha\mu}g_{\beta\nu}-g_{\mu\nu}g_{\alpha\beta})\nabla^{\alpha\beta}f'\bigg]
\label{eq:curv en-mom}
\end{equation}
refers to as an \textit{effective curvature} energy momentum tensor. The former can be interpreted as a sort of curvature fluid characterised by a density
\begin{equation}
\rho_{curv}=\dfrac{1}{f'}\left[\dfrac{1}{2}(f-Rf')-3H\dot{R}f''\right]\,,
\label{eq:rho_curv}
\end{equation}
and a pressure
\begin{equation}
p_{curv}=\dfrac{1}{f'}\left[2H\dot{R}f''+\ddot{R}f''+\dot{R}^2f'''-\dfrac{1}{2}(f-Rf')\right]\,.
\label{eq:p_curv}
\end{equation}
Throughout the text, we use the convention to denote with $'$ the derivative with respect to $R$.

According to the cosmological principle, we assume the spatially flat, homogeneous and isotropic Friedmann-Lema{\^i}tre-Robertson-Walker (FLRW) metric \cite{Planck15}. We thus have $
ds^2= dt^2-a(t)^2\left[dr^2 + r^2( d\theta^2+ \sin^2 \theta\ d\phi^2)\right]$, which implies the constraint over the Ricci scalar and the Hubble parameter:
\begin{equation}
R=-6(\dot{H}+2H^2)\,.
\label{eq:R-H}
\end{equation}
Further, assuming standard matter, i.e. composed by baryons and cold dark matter, with the ansatz that the corresponding EoS is pressureless, we can write
\begin{equation}
H^2=\dfrac{1}{3}\left[\dfrac{1}{f'}\rho_m+\rho_{curv}\right]\,.
\label{eq:Hubble}
\end{equation}
In the Jordan frame, where matter and curvature terms are uncoupled, the continuity equation for the matter fields reads
$
\dot{\rho}_m+3H\rho_m=0
$, which preserves the standard behaviour $
\rho_m=\rho_{m0}a^{-3}=3H_0^2\Omega_{m0}(1+z)^3$, having $\Omega_{m0}$ the value of the matter density today. On the other hand, the continuity equation for the curvature can be written as
\begin{equation}
\dot{\rho}_{curv}+3H(1+w_{curv})\rho_{curv}=3H_0^2\Omega_{m0}(1+z)^3\dfrac{\dot{R}f''}{{(f')}^2}\,,
\label{eq:cont eq curv}
\end{equation}
leading to the curvature EoS:
\begin{equation}\label{eq:w_DE}
w_{DE}\equiv \dfrac{p_{curv}}{\rho_{curv}} =-1+\dfrac{\ddot{R}f''+\dot{R}^2f'''-H\dot{R}f''}{(f-Rf')/2-3H\dot{R}f''}\,,
\end{equation}
which can be supposed to fuel the effective dark energy fluid associated to the curvature.

In this framework, the most relevant caveat remains the difficulty of finding out explicit forms of $f(R)$. The function $f(R)$ is indeed unknown \emph{a priori}. Since different forms of $f(R)$ lead to different cosmological scenarios, it follows that its determination and reconstruction become essential in order to define the correct gravitational theory underlying the universe dynamics \cite{delaCruz16a}. One of the most consolidate approach towards reformulating the form of $f(R)$ is to take data and to frame the $f(R)$ dynamics by means of a back scattering procedure which defines the form of $f(R)$ directly with observations. Unfortunately, Taylor treatments have been so far unsuccessful to determine high-redshift constraints over $f(R)$, spanning from solutions which are either inaccurate as $z\geq1$ or non-univocal. In the next section, we propose to adopt the different expansion due to the Pad\'e approximations in order to alleviate the aforementioned problems.


\section{The method of Pad{\'e} approximants}
\label{sec:Pade}

To overcome the problem of convergence over $f(R)$ Taylor expansions, one can adopt the alternative strategy of expanding series through the use of rational approximants. In this section we present the method of the Pad{\'e} approximants \cite{Baker96}, which is built up from the standard Taylor definition and allows one to reduce divergences at higher redshift domains. In particular, we can express a generic function $f(z)$ as a power series
\begin{equation}
f(z)=\sum_{i=0}^\infty c_iz^i\,,
\label{eq:power series}
\end{equation}
for a given set of coefficients $c_i$. We thus define a $(N,M)$  Pad{\'e} approximant as the ratio
\begin{equation}
P_{N M}(z)=\dfrac{\sum_{n=0}^{N}a_n z^n}{1+\sum_{m=1}^{M}b_m z^m}\,,
\label{eq:def_Padé}
\end{equation}
whose Taylor expansion agrees with \Cref{eq:power series} to the highest possible order, i.e.
\begin{align}
&P_{NM}(0)=f(0)\,,\\
&P_{NM}'(0)=f'(0)\,,\\
&\vdots	\\	
&P_{NM}^{(N+M)}(0)=f^{(N+M)}(0)\,.
\end{align}
The $N+1$ independent coefficients in the numerator and $M$ independent coefficients in the denominator of \Cref{eq:def_Padé} make $N+M+1$ the number of total unknown terms. Hence, we simply can write
\begin{equation}
\sum_{i=0}^\infty c_iz^i=\dfrac{\sum_{n=0}^{N}a_n z^n}{1+\sum_{m=1}^{M}b_m z^m}+\mathcal{O}(z^{N+M+1})\,,
\end{equation}
and, then,
\begin{align}
&(1+b_1z+\hdots +b_Mz^M)(c_0+c_1z+\hdots)= \nonumber \\
&\hspace{1cm}a_0+a_1z+\hdots+a_Nz^N +\mathcal{O}(z^{N+M+1})\,.
\label{coeff}
\end{align}
Equating the terms with the same power coefficients, one obtains a set of $N+M+1$ equations for the $N+M+1$ unknown terms $a_i$ and $b_i$. Depending on the way in which the approximation is built up, for $z \gg 1$, one can use those rational functions as viable candidates to overcome the problems of Taylor series expansions, when one handles high-$z$ data.

The advantage of Pad\'e rational approximations are thus summarized as:

\begin{itemize}
  \item the series can better approximate situations with bad convergence due to data intervals;
  \item the series can easily reduce error bias which propagate when data surveys are not inside $z<1$;
  \item the series can be modeled by choosing appropriate orders which can be chosen depending on each case of interest.
\end{itemize}

Clearly, the Pad\'e polynomials also suffer from precise issues, among them: 

\begin{itemize}
  \item the series convergence is not known a priori, so that the orders should be found by directly comparing with data;
  \item the series can have poles inside the observational domain and this can limit the convergence of the code used to implement data;
  \item different series can degenerate among them, for the net order of Pad\'e series is determined by the sum between the numerator and denominator orders.
\end{itemize}

In what follows, we are interested in assuming the approach of Pad\'e approximations to $f(R)$ gravity and in particular, we need to fix as initial settings over the free-coefficients of the expansions the numerical bounds which can be derived by model-independent measurements inferred from cosmological data. A relevant technique of model-independent reconstruction is offered by cosmography. We thus apply the basic demands of cosmography and the technique of Pad\'e approximation to $f(R)$ gravities.

\section{Cosmographic reconstruction of $f(R)$ cosmology}
\label{sec:cosmography}

Cosmography is a powerful method that allows us to study the present-time cosmology without the need of assuming a specific model to describe the dynamical evolution of the universe. The cosmographic method lies only on the validity of the cosmological principle. Indeed, this model-independent technique does not depend on the solution of the cosmic equations \cite{cosmography1,cosmography2}.
The standard procedure is to expand the scale factor $a\equiv1/(1+z)$ into a Taylor series around the present cosmic time $t_0$:
\begin{equation}
a(t)=1+\sum_{k=1}^{\infty}\dfrac{1}{k!}\dfrac{d^k a}{dt^k}\bigg | _{t=t_0}(t-t_0)^k\,.
\label{eq:scale factor}
\end{equation}
Using \Cref{eq:scale factor}, it is possible to define the so-called \textit{cosmographic parameters}, which represent model-independent quantities that can be directly constrained by observations \cite{Weinberg72,Harrison76,Visser,Poplawski,Luongo13}:
\begin{align}
&H\equiv \dfrac{1}{a}\dfrac{da}{dt} \,, \hspace{1cm} q\equiv -\dfrac{1}{aH^2}\dfrac{d^2a}{dt^2}  \label{eq:H&q} \\
&j \equiv \dfrac{1}{aH^3}\dfrac{d^3a}{dt^3} \,, \hspace{0.5cm}  s\equiv\dfrac{1}{aH^4}\dfrac{d^4a}{dt^4}\,.    \label{eq:j&s}
\end{align}
These quantities are known as\textit{ Hubble rate}, \textit{deceleration parameter}, \textit{jerk} and \textit{snap} parameters, respectively \cite{CS1,CS2}.
From the definition of the luminosity distance
\begin{equation}
d_L(z)=(1+z)\int_0^z\dfrac{dz'}{H(z')}\,,
\label{eq:dL}
\end{equation}
one can use \Cref{eq:scale factor} to obtain the Taylor expansion of the Hubble rate in terms of the cosmographic parameters \cite{Cattoen07}:
\begin{equation}
H(z)=H_0\left(1 + \sum_{\ell=1}^{\infty}\dfrac{1}{\ell!}\dfrac{d^\ell H}{dz^\ell}\bigg | _{z=0} z^\ell \right)\,,
\label{eq:Taylor H}
\end{equation}
the first three orders being
\begin{align}
H_z\big | _{z=0}& =1 + q_0\,,\nonumber\\
H_{zz}\big | _{z=0}&=j_0 - q_0^2\,,\\
H_{zzz}\big | _{z=0}&=\dfrac{1}{6}\Big(j_0(3 +4q_0)-3q_0(1+q_0)+s_0\Big)\,.\nonumber
\end{align}
Here, the subscripts `$z$' denote the derivatives with respect to the redshift.
The impossibility to consider the infinite number of terms of the Taylor polynomials, which would reproduce exactly the real function, leads to truncate the series at some finite order, which is a source of errors in the analysis. Moreover, the Taylor series converges only for small $z$ and it may be inaccurate for analysing high-redshift data. A possible solution to the convergence problem is to consider the Pad{\'e} approximants.  Thus, motivated by the studies done in \cite{Aviles14}, we decide to consider the (2,1) Pad{\'e} approximant of the Hubble rate:
\begin{widetext}
\begin{align}
H_{21}(z)=\ &\Big[2H_0(1 + z)^2 \big(3 + z + j_0 z - q_0 (3 + z + 3 q_0 z)\big)^2 \Big]\times \Big[18 (q_0 - 1)^2 + 6 (q_0 - 1) \big(-5 - 2 j_0 + q_0 (8 + 3 q_0)\big) z \nonumber \\
&+\Big(14 + 2 j_0^2 + j_0 \big(7 - q_0 (10 + 9 q_0)\big) +   q_0 \big(-40 + q_0 (17 + 9 q_0 (2 + q_0))\big)\Big) z^2\Big]^{-1}\,.
\label{eq:H21}
\end{align}
\end{widetext}
We will be comparing the results obtained using \Cref{eq:H21} with the ones deriving from the correspondent standard third-order Taylor expansion for $H(z)$:
\begin{widetext}
\begin{equation}
H(z)\simeq H_0 \left[1 + z (1 + q_0) + \dfrac{z^2}{2} (j_0 - q_0^2) + \dfrac{z^3}{6} \left(-3 q_0^2 - 3 q_0^3 + j_0 (3 + 4 q_0) + s_0\right)\right]\,.
\label{eq:H_Taylor}
\end{equation}
\end{widetext}
It is worth noting that the  $H_{21}(z)$ contains the cosmographic parameters up to the jerk, while in the case of third-order Taylor approximation also the snap is present.

A common procedure for studying $f(R)$ gravity is to assume a specific $f(R)$ function and solve the modified Friedmann equation to obtain $H(z)$. However, this method relies on the \textit{a priori} choice of $f(R)$ and, thus, on the assumption of the cosmological model. Here, we show a method that allows us to reconstruct the function $f(R(z))=f(z)$ in a model-independent way. 
 Cosmography can be used in the framework of $f(R)$ gravity and in general in the field of any modified theory for several reasons. In fact, once the definitions of the cosmographic parameters are provided, it is possible to relate the form of $f(R)$ in terms of this set by inverting the modified Friedmann equations. This process is \emph{exact}, \textit{i.e.}  does not need the Taylor approximation of $a(t)$. In such a way, once the cosmographic series is known by direct comparison with data, in principle one can go further with $f(R)$ to get bounds over $f(R)$ and its derivatives. Moreover, it could be also possible to frame the $f(R)$ evolution if the cosmographic series was known at all redshifts $z$. Unfortunately, in the framework of $f(R)$ gravity, inverting the Friedmann equation is only possible numerically. Moreover, the present status of cosmography defines today only background results\footnote{A high redshift cosmography would be a future goal to understand how the universe evolves in a model-independent way.},\textit{ i.e.} bounded at $z\simeq0$.  Applications of cosmography to get constraints over $f(R)$ and $f(T)$ have been properly investigated in the literature, e.g. for example \cite{sa77}.
In all these approaches, one recovers the above motivations. The authors considered cosmography as a way to rewrite cosmic quantities of interest and to enable a much quicker inversion of each term entering the modified Friedmann equations. In our work, we consider the Pad\'e expansions to guarantee convergence over the cosmographic set at higher redshift, \textit{i.e.} to enable the cosmographic predictions in a redshift domain much larger than the standard one predicted by Taylor series.

\begin{figure}[h!]
\begin{center}
\includegraphics[width=3.2in]{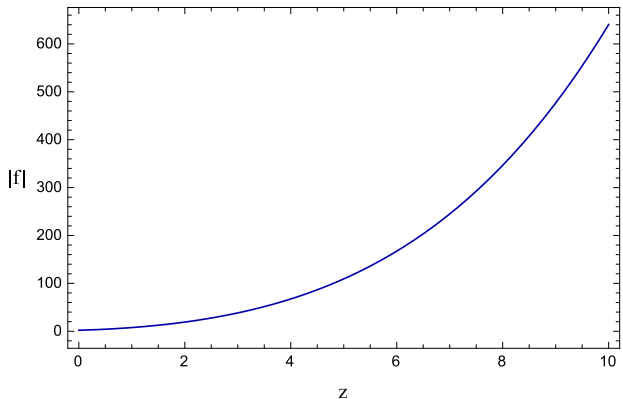}
\caption{Numerical reconstruction of $|f(z)|$ using the (2,1) Pad{\'e} approximant.}
\label{fig:f(z) Padé}
\end{center}
\end{figure}

Our strategy consists of combining \Cref{eq:Hubble} and \Cref{eq:rho_curv} and using \Cref{eq:R-H}, once the values of the cosmographic parameters are known. To do this, we need to express the derivatives with respect to time and with respect to $R$ as derivatives with respect to $z$. Being $F(z)$ an arbitrary function depending on the redshift, one has
\begin{align}
\dfrac{dF}{dt}&=-(1+z)HF_z\,,	\label{eq:dt-dz}	\\
\,\nonumber\\
\dfrac{\partial F}{\partial R}&=\dfrac{1}{6} \Big[(1 + z)H_z^2 +H\left(-3H_z +(1+z)H_{zz}\right)\Big]^{-1}F_z\,.  \label{eq:dR-dz}
\end{align}
 This leads to a second-order differential equation for $f(z)$:
\begin{widetext}
\begin{align}
H^2f_z&=\Big[-(1+z)H_z^2+H\big(3H_z-(1+z)H_{zz}\big)\Big]\Bigg[-6 H_0^2 (1 + z)^3 \Omega_{m0} - f -\dfrac{Hf_z \left(2 H - (1 + z) H_z\right)}{(1 + z)H_z^2 +H\left(-3H_z+ (1 + z)H_{zz}^2\right)} \nonumber \\
&-\dfrac{(1 + z) H^2\Big(f_{zz}\big((1 +z)H_z^2 + H (-3H_z + (1 + z)H_{zz})\big)+f_z\big(2H_z^2-3(1+z)H_zH_{zz}+H(2H_{zz}-(1+z)H_{zzz})\big)\Big) }{{\big[(1+z)H_z^2+H\big(-3H_z+(1+z)H_{zz}\big)\big]}^2} \Bigg]\,.
\label{eq:f(z)}
\end{align}
\end{widetext}
\Cref{eq:f(z)} requires two initial conditions over $f$ and $f_z$ to be solved. 
These can be obtained by means of \Cref{eq:rho_curv,eq:p_curv,eq:Hubble} evaluated at the present time, together with the condition $f'(R_0)=1$ which guaranties that the effective gravitational constant of the theory, $G_{eff} = G_N/f'(R)$, matches the Newtonian constant $G_N$ today. One, thus, finds:
\begin{align}
&f_0=R_0+6H_0^2(\Omega_{m0}-1)\ , \label{eq:f0}\\
&f_z\big|_{z=0}=R_z\big|_{z=0}\ . \label{fp0}
\end{align}
Throughout our analysis, we fix $\Omega_{m0}=0.3$. As far as the cosmographic parameters are concerned, we use the results found in \cite{Aviles14}. For the (2,1) Pad{\'e} approximant, these read
\begin{equation}
\left\{
\begin{aligned}
&h=0.7064^{+0.0277}_{-0.0263}\,, \\
&q_0=-0.4712^{+0.1224}_{-0.1106}\,,\\
&j_0=0.593^{+0.216}_{-0.210}\,,\\
\end{aligned}
\right .
\label{eq:cosm param Padé}
\end{equation}
while, in the case of the third-order Taylor expansion, we have
\begin{equation}
\left\{
\begin{aligned}
&h=0.7253^{+0.0353}_{-0.0351}\,,\\
&q_0=-0.6642^{+0.2050}_{-0.1963}\,,\\
&j_0=1.223^{+0.644}_{-0.664}\,,\\
&s_0=0.394^{+1.335}_{-0.731}\,,
\end{aligned}
\right .
\label{eq:cosmogr param Taylor}
\end{equation}
where $h\equiv H_0/(100\ \text{km/s/Mpc})$.
Plugging \Cref{eq:H21} into \Cref{eq:f(z)}, we are able to reconstruct $f(z)$ numerically. 
Due to the negative sign of $R$ as a consequence of the metric signature we have adopted, our reconstructed $f(R)$ function will be negative. This, in turns, implies that $f(z)$ must to be negative, which is consistent with choosing the upper bound values of \Cref{eq:cosm param Padé}. 
\Cref{fig:f(z) Padé} shows the numerical results for the (2,1) Pad{\'e} approximant.

The following test-functions have been considered in order to find an analytical form of $f(z)$ that matches the numerical results:
\begin{subequations}
\begin{align}
Exponential\nonumber\\
f_1(z)&=\mathcal A z+\mathcal B z^3 e^{\mathcal C z}	               \label{eq:exp}\\
f_2(z)&=\mathcal A+\mathcal B z^2 \sinh(1+\mathcal C z)				\label{eq:sinh}\\
f_3(z)&=\mathcal A z+\mathcal B z^3 \cosh(\mathcal C z)		        \label{eq:cosh}\\
f_4(z)&=\mathcal Az^2+\mathcal B z^4\tanh(\mathcal C z)	            \label{eq:tanh}\\
Trigonometric\nonumber\\
f_5(z)&=\mathcal Az^3+\mathcal B z^5 \sin(1+\mathcal C z)         \label{eq:sin}\\
f_6(z)&=\mathcal Az^3+\mathcal B z^4 \cos(1+ \mathcal C z)                           \label{eq:cos}\\
f_7(z)&=\mathcal Az+\mathcal B z^2 \tan(\mathcal C z)	                \label{eq:tan}\\
Logarithmic\nonumber\\
f_8(z)&=\mathcal Az+\mathcal B z^3 \ln(1+\mathcal C z) 		       \label{eq:ln}
\end{align}
\end{subequations}
where the set of three constants, $\mathcal{A}$, $\mathcal{B}$ and $\mathcal{C}$, includes free parameters determined through a fitting procedure.

\subsection{Statistics and strategy of data mining}

\begin{figure}[h!]
\begin{center}
\includegraphics[width=3.1in]{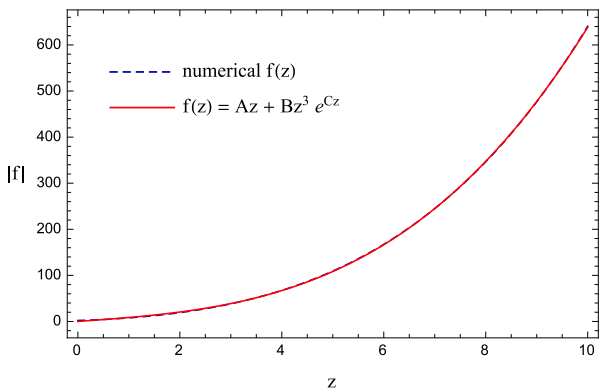}
\caption{Comparison between the numerical reconstruction of $|f(z)|$ and the functional form \Cref{eq:approx f} for the (2,1) Pad{\'e} approximant.}
\label{fig:approx f(z)}
\end{center}
\end{figure}

In order to find the best analytical approximation for our models, we perform the $\mathcal{F}$-\textit{statistics} \cite{James13}, defined by:
\begin{equation}
\mathcal F=\dfrac{(\text{TSS}-\text{RSS})/p}{\text{RSS}/(n-p-1)}\,,
\label{eq:F-statistics}
\end{equation}
where
\begin{align}
&\text{TSS}=\sum_{i=1}^{n}(y_i-\bar{y})^2\,, \\
&\text{RSS}=\sum_{i=1}^n(y_i-\hat{y}_i)^2\,,
\end{align}
are the\textit{ total sum of squares} and the \textit{residual sum of squares}, respectively. Here,
\begin{equation}
\bar{y}=\dfrac{1}{n}\sum_{i=1}^n y_i\,,
\end{equation}
$y_i$ represents the $i$-th observed response value and $\hat{y}_i$ the $i$-th response value predicted by the model, while $n$ is the number of observations and $p$ the number of predictors. The $\mathcal{F}$-\textit{statistics} provides a measure of the goodness of a model by testing the joint explanatory power of its predictors. The null hypothesis, i.e. that all the $p$ regression coefficients are zero and the model has no explanatory power, is tested against the alternative hypothesis that at least one of the regression coefficients is different from zero.  The $\mathcal{F}$-\textit{statistics} presents some advantages compared with tests that look for any association between the individual variables and the response, such as $t$-\textit{statistics} and $p$-\textit{value}, or compared with $\mathcal{R}^2$-\textit{test} since it adjusts the number of predictors. In fact, when $p$ is large, there is a very high chance to observe small $p$-\textit{values} even in absence of any real association between the predictors and the response. Further, the $\mathcal{R}^2$-\textit{test} may be often misleading as $\mathcal{R}^2$ always increases when more predictors are added to the model, even if those variables are only weakly associated with the response. The $\mathcal{F}$-\textit{statistics} can also be expressed in terms of $\mathcal{R}^2$ as
\begin{equation}
\mathcal{F}=\dfrac{\mathcal{R}^2/p}{(1-\mathcal{R}^2)/(n-p-1)}\,.
\end{equation}
If the null hypothesis is true, we expect $\mathcal{R}^2$ and $\mathcal{F}$ to be close to zero. Thus a high value of the $\mathcal{F}$-\textit{statistics} indicates evidence for the model against the null hypothesis.
\begin{table}[h!]
\begin{center}
\begin{tabular}{|c|c|c|}
\hline
Test-function & $(\mathcal A, \mathcal B, \mathcal C)$ & $\mathcal F (\times 10^6)$  \\
\hline
$f_1(z)$ & $(-8.078, -0.530, 0.005)$  &  $31.7 $ \\
$f_2(z)$ & $( -6.147, -2.148, 0.080)$ & $13.5$  \\
$f_3(z)$ & $(-8.046, -0.541, 0.025)$ & $3.637$  \\
$f_4(z)$ & $(-3.699,  0.027,  -562.2)$& $4.535$  \\
$f_5(z)$ & $(-0.708,  -0.001, 1.095)$ & $0.118$   \\
$f_6(z)$ & $(-0.717, -0.008,  0.)$ & $0.142$  \\
$f_7(z)$ & $(-41.30, 0.002, 1.000)$ & $0.026$   \\
$f_8(z)$ & $(-11.69, -0.208, 1.182)$ & $1.484$\\
\hline
\end{tabular}
\caption{$\mathcal{F}$-\textit{statistics} on the test-functions \Crefrange{eq:exp}{eq:ln} for the (2,1) Pad{\'e} approximant.}
\label{tab:F-Padé}
\end{center}
\end{table}
In our case, $p=3$ and we generate $n=1000$ points from the numerical solution of $f(z)$.  As shown in \Cref{tab:F-Padé}, the best analytical match to the numerical $f(z)$ for the Pad{\'e} approximant results to be
\begin{equation}
f(z)=\mathcal A z+\mathcal B z^3 e^{\mathcal C z}\,,
\label{eq:approx f}
\end{equation}
with
\begin{equation}
(\mathcal A, \mathcal B, \mathcal C)=(-8.078, -0.530, 0.005)\,.
\label{eq:coeff 1}
\end{equation}
\Cref{fig:approx f(z)} shows the comparison between the numerical solution of $f(z)$ and the analytical function \Cref{eq:approx f} for the Pad{\'e} approximant in the redshift domain $z\in[0,10]$.

\section{Cosmological consequences}
\label{sec:cosmology}

\begin{figure}[h!]
\begin{center}
\includegraphics[width=3.2in]{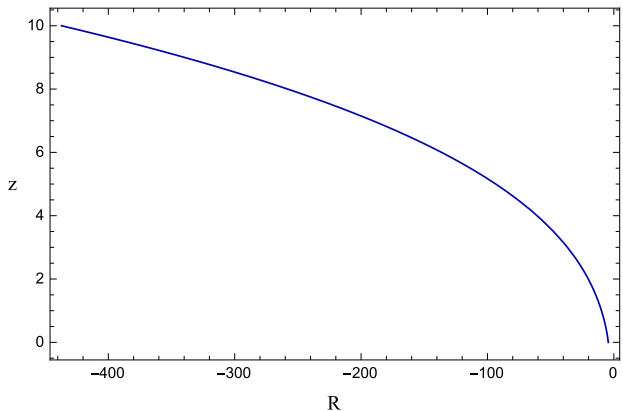}
\caption{Numerical solution of $z(R)$ for the (2,1) Pad{\'e} approximant.}
\label{fig:z(R)}
\end{center}
\end{figure}

\begin{figure}[h!]
\begin{center}
\includegraphics[width=3.2in]{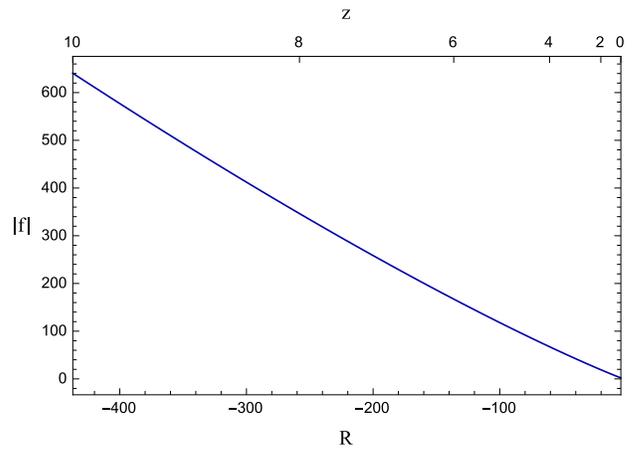}
\caption{Reconstructed $|f(R)|$ for the (2,1) Pad{\'e} approximant in the redshift domain $z\in[0,10]$.}
\label{fig:f(R)}
\end{center}
\end{figure}

To obtain $f(R)$, we need to find $R$ as a function of the redshift $z$ and then to invert it to have $z(R)$, which will be plugged back into \Cref{eq:approx f}. Unfortunately, \Cref{eq:H21} cannot be inverted analytically, which drives us to use the numerical result. Thus, by means of \Cref{eq:R-H}, we are able to find $z(R)$ (see \Cref{fig:z(R)}).
This can be inserted into \Cref{eq:approx f} to finally get $f(R)$ (see \Cref{fig:f(R)}).

\subsection{Viability conditions for $f(R)$ models}

For a viable explanation to dark energy, $f(R)$ models have to satisfy certain conditions. In the context of the metric formalism, the first condition is
\begin{equation}
f'(R)>0\ , \hspace{0.2cm} R\geq R_0\ 
\label{condition f'(R)}
\end{equation}
if $R_0>0$. This condition is required in order to avoid negative values of the effective gravitational constant. 
Then, the second condition reads
\begin{equation}
f''(R)>0\ , \hspace{0.2cm} R\geq R_0
\label{condition f''(R)}
\end{equation}
if $R_0>0$. This arises from the constraints of tests of gravity in the solar system \cite{solar system}, and the consistency with the presence of a standard matter-dominated epoch \cite{Amendola07}. Moreover, this condition guaranties the stability of cosmological perturbations \cite{cosm perturb}.

\begin{figure}[h!]
\begin{center}
\includegraphics[width=3.2in]{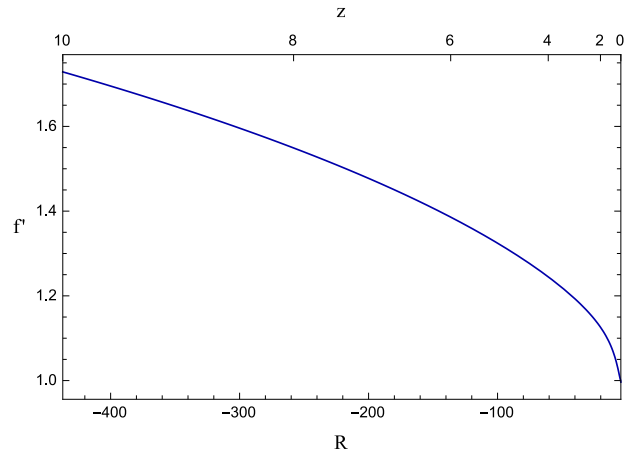}
\caption{Functional behaviour of $df/dR$ as result of the (2,1) Pad{\'e} approximant in the redshift domain $z\in[0,10]$.}
\label{fig:f'(R)}
\end{center}
\end{figure}

\begin{figure}[h!]
\begin{center}
\includegraphics[width=3.3in]{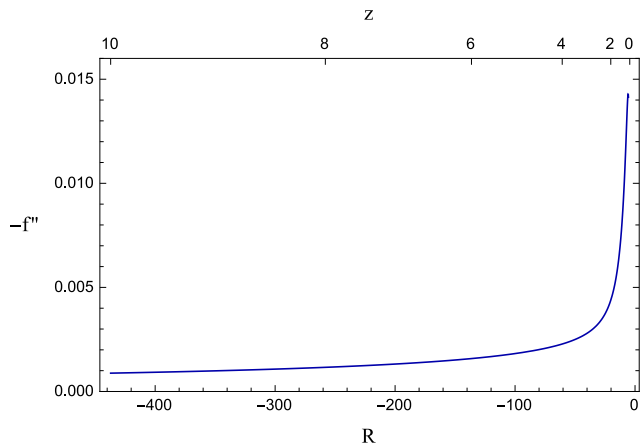}
\caption{Functional behaviour of $d^2f/dR^2$ as result of the (2,1) Pad{\'e} approximant in the redshift domain $z\in[0,10]$. We show $-f''(R)$ for display purposes.}
\label{fig:f''(R)}
\end{center}
\end{figure}

To verify whether our model fullfils the above conditions, we display $f'(R)$ and $f''(R)$ in \Cref{fig:f'(R)} and \Cref{fig:f''(R)}, respectively.
We note that both conditions are satisfied, which ensures the cosmological viability of our model.

\begin{figure}[h!]
\begin{center}
\includegraphics[width=3.2in]{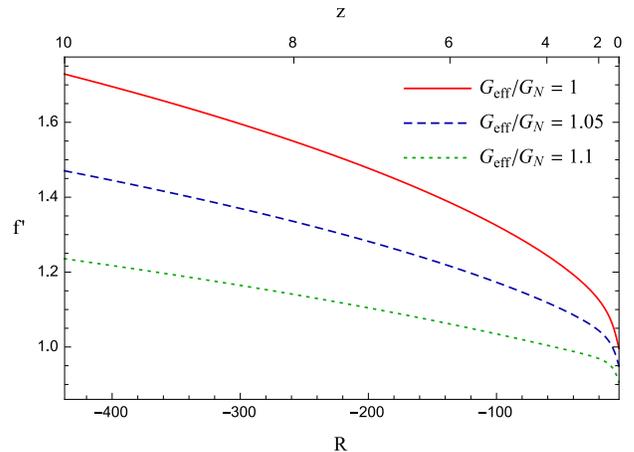}
\caption{Functional behaviour of $f'(R)$ for different values of the effective gravitational constant.}
\label{fig:Geff}
\end{center}
\end{figure}

Constraints obtained from the Cosmic Microwave Background require  that a viable $f(R)$ model approaches to $\Lambda$CDM for large curvatures. To ensure this behavior, one expects to fulfill the requirement that $f'(R)\rightarrow1$ for $R\gg 1$ \cite{lcdm limit}. However, \Cref{fig:f'(R)} indicates that $f'(R)$ slightly exceeds unity in the limit of large curvatures. This is due to the fact that the asymptotic value of $f'(R)$ depends on the the initial settings adopted for $f'(R_0)$. To demonstrate this fact, one can weakly relax the assumption $f'(R_0)=1$, allowing small departures from the Newtonian gravitational constant, i.e. requiring that $G_{eff}$ is not exactly equivalent to the Newton  $G$. This choice does not violate the limits on $\dot G/G$ as imposed in current literature \cite{Martins17}.
We thus infer that \Cref{eq:f0,fp0} take the following expressions:
\begin{align}
&f_0=f'(R_0)(6 H_0^2 + R_0) - 6 H_0^2 \Omega_{m0}\ , \\
&f_z\big|_{z=0}=f'(R_0)\ R_z\big|_{z=0}\ .
\end{align}
Indeed, using the above relations as initial values to get the auxiliary function $f(z)$, we soon obtain the results displayed in \Cref{fig:Geff}.
Last but not least, a further factor which affects the asymptotic value of $f'(R)$ is related to the determination of the cosmographic series. Bearing in mind that the predictive power of the cosmographic method degrades as the redshift increases, one should consider higher-order Pad\'e polynomials to improve the convergence radii of the cosmographic series.
This issue is also known as \emph{cosmographic convergence problem} \cite{Dunsby16} and clearly influences any treatments at high-redshift domains. By adopting the aforementioned settings, the difference $f'(R)_{numerical}-f'(R)_{exact}$ is small at larger curvatures and it is clearly due to the approximations made on the orders, initial values, and convergence of the Pad\'e polynomials.

\subsection{Dark energy equation of state}

The reconstructed $f(R)$ we have obtained can be used to find $\rho_{curv}$ and $p_{curv}$ as functions of the Ricci scalar and, therefore, to study the dark energy EoS, i.e. $w_{DE}$ (cf. \Cref{eq:w_DE}).
To improve the error propagation in our numerical analyses, we re-scale \Cref{eq:approx f} as follows
\begin{equation}
f(z)\longrightarrow \lambda+f(z)\,.
\label{eq:rescale}
\end{equation}
Here, the parameter $\lambda$ means  that we are using test-functions to numerically reconstruct the form of $f(z)$. Indeed, its role is to tune the numerical result and to enable a Taylor expansion over the expressions for $f(z)$, here involved for understanding the evolution of $f(R)$. To figure this out, it is simple to check that it does not come as vacuum energy contribution since it acts as a scaling constant to guarantee that at $z=0$ the value of $f(z)$ is always compatible with the fact that $f'(R_0)=1$. Moreover, its magnitude is ten times higher than the critical density, being outer the limit which enables $H(z=0)=H_0$ and then cannot be considered as a vacuum energy entering the weak energy condition. 

\begin{figure}[h!]
\begin{center}
\includegraphics[width=3.2in]{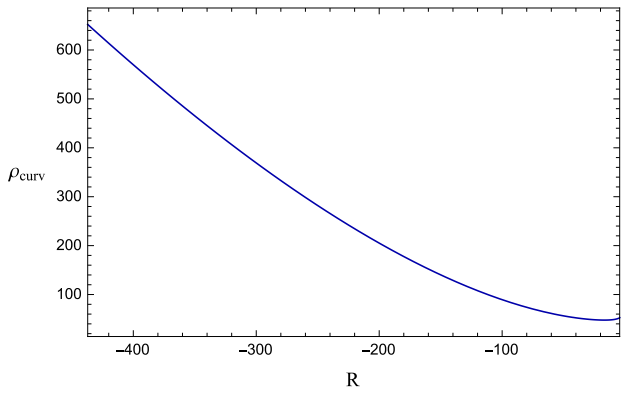}
\caption{Effective curvature density for the Pad{\'e} approximant with $\lambda=100$.}
\label{fig:rho(R)}
\end{center}
\end{figure}

\begin{figure}[h!]
\begin{center}
\includegraphics[width=3.2in]{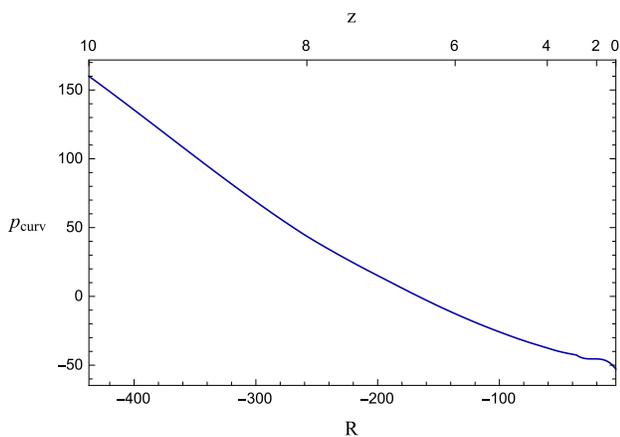}
\caption{Effective curvature pressure for the Pad{\'e} approximant with $\lambda=100$.}
\label{fig:p(R)}
\end{center}
\end{figure}

\begin{figure}[h!]
\begin{center}
\includegraphics[width=3.3in]{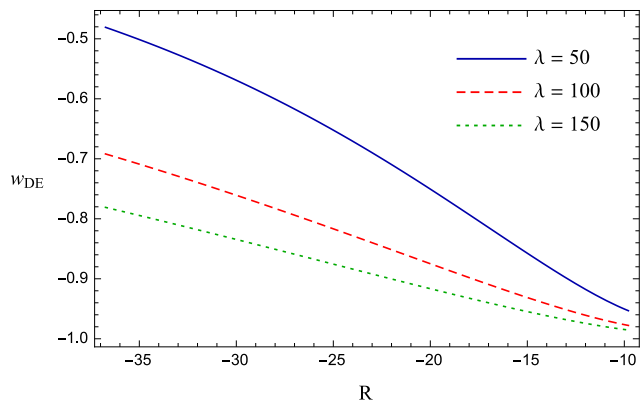}
\caption{Effective dark energy EoS parameter for the Pad{\'e} approximant for different values of the constant $\lambda$.}
\label{fig:w(R)}
\end{center}
\end{figure}

The constant $\lambda$ is found requiring the accelerated universe today, i.e.
\begin{equation}
-1\leq w_{DE}\Big|_{z=0}<-\dfrac{1}{3}\,,
\label{eq:constraint on w}
\end{equation}
getting the constraint
\begin{equation}
\lambda \gtrsim 19.3\,.
\label{eq:constraint on lambda}
\end{equation}
\Cref{fig:rho(R)} and \Cref{fig:p(R)} show the behaviours of curvature density and pressure for an indicative value of $\lambda=100$.
In \Cref{fig:w(R)}, we show the effective dark energy EoS parameter for different outcomes of $\lambda$, satisfying the condition \ref{eq:constraint on lambda}.

\section{Taylor expansion vs Pad{\'e} approximation}
\label{sec:Taylor}

\begin{figure}[h!]
\begin{center}
\includegraphics[width=3.2in]{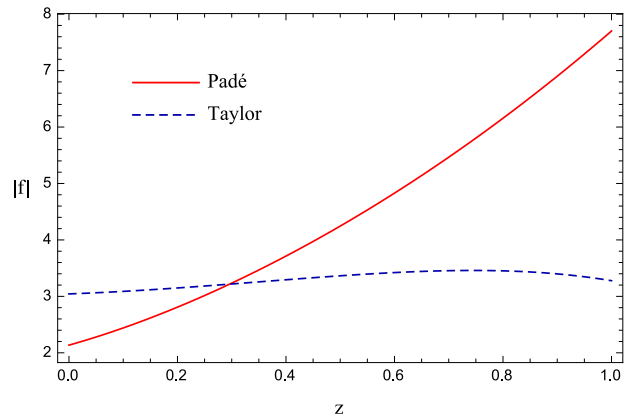}
\caption{Comparison between the numerical reconstruction of $|f(z)|$ using the (2,1) Pad{\'e} (solid red) and the third-order Taylor (dashed blue) approximations.}
\label{fig:f(z) Pade vs Taylor}
\end{center}
\end{figure}

\begin{figure}[h!]
\begin{center}
\includegraphics[width=3.2in]{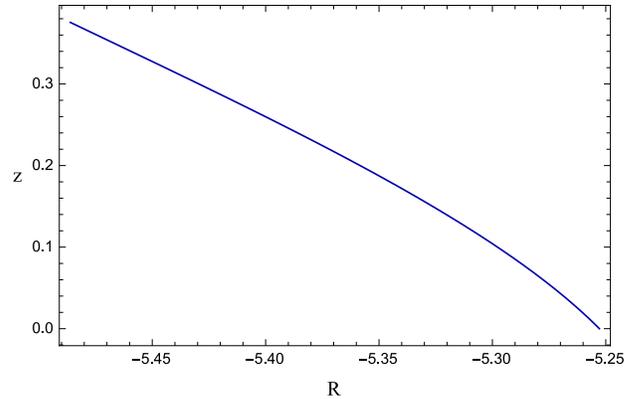}
\caption{Numerical solution of $z(R)$ for the third-order Taylor approximation.}
\label{fig:z(R) Taylor}
\end{center}
\end{figure}

\begin{figure}[h!]
\begin{center}
\includegraphics[width=3.2in]{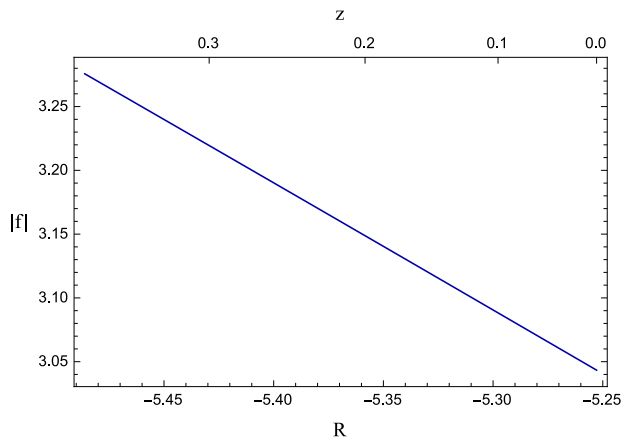}
\caption{Reconstructed $f(R)$ for the third-order Taylor approximation.}
\label{fig:f(R) Taylor}
\end{center}
\end{figure}

To better check the benefits of our analysis, based on Pad{\'e} approximations with respect to the standard series approach, we present the results one would obtain using the Taylor method. Using \Cref{eq:H_Taylor}, we can solve \Cref{eq:f(z)} by adopting the best-fit results of \ref{eq:cosmogr param Taylor}. \Cref{fig:f(z) Pade vs Taylor} shows the comparison between the best-fit results based on the Pad{\'e} and the Taylor approximations in the redshift interval $z\in[0,1]$. We can see that the Taylor approximation stops being predictive already at $z\simeq 0.3$. At higher redshifts, the two approaches have very different behaviours.
The numerical inversion of \Cref{eq:H_Taylor} by means of \Cref{eq:R-H} yields $z(R)$ (see \Cref{fig:z(R) Taylor}), which inserted back in $f(z)$, allows us to find the form of $f(R)$ for the Taylor approximation (see \Cref{fig:f(R) Taylor}).

\begin{figure}[h!]
\begin{center}
\includegraphics[width=3.3in]{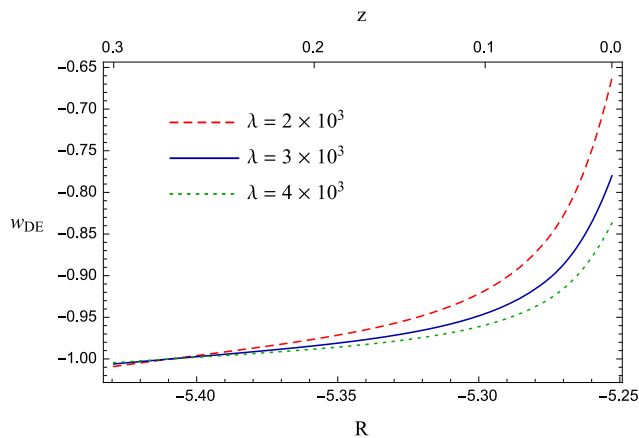}
\caption{Effective dark energy EoS parameter for the third-order Taylor approximation for different values of the constant $\lambda$.}
\label{fig:w_Taylor}
\end{center}
\end{figure}

We shall study the dark energy EoS parameter (cf. \Cref{eq:w_DE}) using the Taylor approach and compare the results with those we have obtained for the Pad{\'e} approximation. In the case of the Tayor approximation, the rescaling \ref{eq:scale factor} together with the condition \ref{eq:rescale} lead to
\begin{equation}
\lambda \gtrsim 1196\,.
\label{eq:constraint on lambda 2}
\end{equation}
As shown in \Cref{fig:w_Taylor}, the dark energy EoS parameter crosses the phantom line ($w_{DE}=-1$) at $z\sim 0.3$, confirming the inability of the Taylor method to account for observations at higher redshifts.

\section{Conclusions and perspectives}
\label{sec:conclusions}

The $f(R)$ gravity models have been revised here, considering a strategy to reconstruct at high redshift the functional forms of $f(R)$. In particular, we presented a technique for reconstructing the form of $f(R)$ in a model-independent way, without resorting to any \emph{a priori} assumptions. To do so, we applied the cosmographic method with the use of Pad{\'e} rational polynomials, showing the advantages of this treatment with respect to the standard method based on Taylor expansions, especially when the redshift domain exceeds $z\simeq1$. To figure this out, expanding the Hubble rate $H(z)$ in terms of the cosmographic series and using the relation $R=-6(\dot{H}+2H^2)$ in a FRW universe, we found  $f(z)\equiv f(R(H))$ by numerically solving the modified Friedmann equations. We compared the results obtained for a (2,1) Pad{\'e} approximant, involving up to the jerk parameter, with the outcomes of the third-order Taylor approach, when also the snap comes into the analysis. The $\mathcal{F}$-\textit{statistics} applied to several test-functions showed that the most suitable choice for $f(z)$ is the form $f(z)=\mathcal A z+\mathcal B z^3 e^{\mathcal C z}$, where the free parameters $\mathcal A$, $\mathcal B$, and $\mathcal C$ were found in order to match the cosmographic parameters to the values suggested by the most recent observations. Then, through a back-scattering procedure, we inverted the constraint $R=-6(\dot{H}+2H^2)$ to find  $z(R)$ and we finally reconstruct the term $f(R)$.
We showed that the so-obtained $f(R)$ model satisfies all the conditions required from solar system tests of gravity and cosmological perturbations theory and, therefore, it represents a viable model to explain dark energy.
The cosmological implications of our model have been investigated by studying the EoS of the effective dark energy fluid in the redshift interval $z\in[0,10]$. We performed statistical analyses based on basics demands of data mining, employing in particular the $\mathcal F$ test and other strategies which overcome problems related to the $p$-value analysis and $\chi^2$ square procedure. We thus selected the effective dark energy behaviour and we reproduced its functional evolution at both the small and high redshift domains. 
By increasing the approximation order of the Pad\'e series and slightly changing the initial settings on the differential equation, it is possible to reduce the discrepancy between $f'(R)_{numerical}$ and $f'(R)_{exact}$ in order to show that the condition $f'(R)\rightarrow 1$ for $R\rightarrow\infty$ is naturally guaranteed.
Future analyses will also involve tests on the Cosmic Microwave Background in order to check the accuracy of our numerics.

\section*{Acknowledgments}
S.C. acknowledges the support of INFN (iniziativa specifica QGSKY). This paper is based upon work from COST action CA15117 (CANTATA), supported by COST (European Cooperation in Science and Technology).


\begin{thebibliography}{99}

\bibitem{Copeland06}
E. J. Copeland, M. Sami, S. Tsujikawa, Int. J. Mod. Phys. D, {\bf 15}, 1753-1936, (2006).

\bibitem{Bamba12}
K. Bamba, S. Capozziello, S. Nojiri, S. D. Odintsov, Astrophys. Space Sci., {\bf 342}, 155 -228, (2012).

\bibitem{Davis16}
T. M. Davis, D. Parkinson, Handb. Super., {\bf 1}, (2016).

\bibitem{Joyce16}
A. Joyce, L. Lombriser, F. Schmidt, Annu. Rev. Nucl. Part. Sci., {\bf 66}, 95-122, (2016).

\bibitem{Kleidis16}
K. Kleidis, N. K. Spyrou, Entropy, {\bf 18}, 3-94, (2016).

\bibitem{Carroll92}
S. M. Carroll, W.H. Press, E.L. Turner, ARAA, {\bf 30}, 499, (1992).

\bibitem{Sahni00}
V. Sahni, A. Starobinski, Int. J. Mod. Phys. D, {\bf 9}, 373, (2000).

\bibitem{SNe}
S. Perlmutter et al., Nature, {\bf 391}, 51-54, (1998);
B. P. Schmidt et al., Astrophys. J., {\bf 507}, 46-63, (1998);
A. G. Riess et al., Astron. J., {\bf 116},1009-1038, (1998).

\bibitem{WMAP9}
G. Hinshaw et al. [WMAP Collaboration], Astrophys. J. Suppl., {\bf 208}, 19, (2013).

\bibitem{Planck15}
P. A. R. Ade et al. [Planck Collaboration],  Astron. Astrophys, {\bf 594}, A13, (2016).

\bibitem{cosmography1}
T. D. Saini, S. Raychaudhury, V. Sahni, A. A. Starobinsky, Phys. Rev. Lett., {\bf 85}, 1162, (2000);
O. Luongo, Mod. Phys. Lett. A, {\bf 26}, 20, 1459, (2011).

\bibitem{cosmography2}
J. C. Carvalho, J. S. Alcaniz, Mon. Not. Roy. Astron. Soc., {\bf 418}, 1873-1877, (2011);
L. Xu, Y. Wang, Phys. Lett. B, {\bf 702}, 114-120, (2011);
A. Aviles, C. Gruber, O. Luongo, H. Quevedo, Phys. Rev. D, {\bf 86}, 123516, (2012).

\bibitem{Weinberg89}
S. Weinberg, Rev. Mod. Phys., {\bf 61},1-23, (1989).

\bibitem{Zlatev99}
I. Zlatev, Li-Min Wang, P. J. Steinhardt, Phys. Rev. Lett., {\bf 82}, 896-899, (1999).

\bibitem{Sahni02}
V. Sahni, Class. Quant. Grav., {\bf 19}, 3435-3448, (2002).

\bibitem{Sotiriou10}
T. P. Sotiriou, V. Faraoni, Rev. Mod. Phys., {\bf 82}, 451, (2010).

\bibitem{De Felice10}
A. De Felice, S. Tsujikawa, Living Rev. Rel. {\bf 13}, 3, (2010).

\bibitem{Odintsov}
S. Nojiri, S. D. Odintsov, Phys. Rept. \textbf{505},  59-144 (2011);
S. Nojiri, S. D. Odintsov, V. K. Oikonomou, Phys. Rept. \textbf{692},  1-104 (2017).

\bibitem{Capozziello11}
S. Capozziello, M. De Laurentis, Phys. Rept., {\bf 509}, 167-321, (2011).

\bibitem{Hu07}
W. Hu, I. Sawicki, Phys. Rev. D, {\bf 76}, 064004, (2007).

\bibitem{Starobinsky07}
A. A. Starobinsky, JETP Lett. {\bf 86}, 157, (2007).

\bibitem{Appleby07}
S. A. Appleby, R. A. Battye, Phys. Lett. B, {\bf 654}, 7, (2007).

\bibitem{Tsujikawa08}
S. Tsujikawa, Phys. Rev. D, {\bf 77}, 023507, (2008).

\bibitem{Cognola08}
G.  Cognola et  al., Phys. Rev. D, {\bf 77}, 046009 (2008).

\bibitem{Gruber13}
C. Gruber, O. Luongo, Phys. Rev. D, {\bf 89}, 103506, (2014).

\bibitem{Capozziello}
S. Capozziello,  Int. J. Mod. Phys. D, {\bf 11}, 483-492, (2002);
S. Capozziello, V. F. Cardone, S. Carloni, A. Troisi, Int. J. Mod. Phys. D, {\bf 12}, 1969-1982, (2003);
S. Capozziello, V. F. Cardone, A. Trosi, Phys. Rev. D, 71, 043503, (2005);
S. Carloni, P. K. S. Dunsby, S. Capozziello, A. Troisi, Class. Quant. Grav., {\bf 22}, 4839, (2005).

\bibitem{altri}
H. Kleinert, H.-J. Schmidt, Gen. Rel. Grav. {\bf 34}, 1295, (2002);
S. M. Carroll, V. Duvvuri, M. Trodden, M. Turner, Phys. Rev. D, {\bf 70}, 043528, (2004);
G. Allemandi, A. Borowiec, M. Francaviglia, Phys. Rev. D, {\bf 70}, 103503, (2004).

\bibitem{delaCruz16a}
A. de la Cruz-Dombriz, P. K. S. Dunsby, S. Kandhai, D. Saez-Gomez, Phys. Rev. D, {\bf 93}, 084016 (2016).

\bibitem{Baker96}
G. A. Baker Jr., P. Graves-Morris, \emph{Pad{\'e} Approximants}, Cambridge University Press, (1996).

\bibitem{Weinberg72}
S. Weinberg, \emph{Gravitation and cosmology}, Wiley, New York, (1972).

\bibitem{Harrison76}
E. Harrison, Nature, {\bf 260}, 591, (1976).

\bibitem{Visser}
M. Visser, Phys. Rev. D, {\bf 56}, 7578, (1997); M. Visser, Gen. Rel. Grav., {\bf 37}, 1541, (2005); M. Visser, Class. Quant. Grav., {\bf 32}, 135007, (2015).

\bibitem{Poplawski}
N. Poplawski, Phys. Lett. B, {\bf 640}, 135, (2006); N. Poplawski, Class. Quant. Grav., {\bf 24}, 3013, (2007).

\bibitem{Luongo13}
O. Luongo, Mod. Phys. Lett. A, {\bf 28}, 1350080, (2013).

\bibitem{CS1}
E.R. Harrison, Nature, {\bf 260}, 591, (1976);
P. T. Landsberg, Nature, {\bf 263}, 217, (1976);
T. Chiba, Prog. Theor. Phys., {\bf 100}, 1077, (1998);
Y. Shtanov, V. Sahni, Class. Quantum Grav. {\bf 19}, L101, (2002);
U. Alam,  V. Sahni, T. D. Saini, A. A. Starobinsky,  Mon. Not. R. Astron. Soc. {\bf 344}, 1057, (2003);
V. Sahni, T.D. Saini, A.A. Starobinsky, U. Alam, JETP Lett. {\bf 77}, 201, (2003).

\bibitem{CS2}
R. R. Caldwell,  M. Kamionkowski, JCAP, {\bf 0409}, 009, (2004);
M. P. Dabrowski, T. Stachowiak, Annals Phys., {\bf 321}, 771-812, (2006).

\bibitem{Cattoen07}
C. Cattoen, M. Visser, Class. Quant. Grav., {\bf 24}, 5985-5998, (2007) .

\bibitem{Aviles14}
A. Aviles, A. Bravetti, S. Capozziello, O. Luongo, Phys. Rev. D, {\bf 90}, 04353, (2014).


\bibitem{sa77}
S. Capozziello, V. F. Cardone, V. Salzano, Phys. Rev. D, {\bf 78}, 063504, (2008); 
S. Capozziello, V. Salzano, Adv. Astron., {\bf 1}, (2009); S. Capozziello, E. De Filippis, V. Salzano, Mon. Not. Roy. Astron. Soc., {\bf 394}, 947, (2009).


\bibitem{James13}
G. James, D. Witten, T. Hastie, R. Tibshirani, \emph{An Introduction to Statistical Learning}, Springer-Verlag New York, (2013).


\bibitem{solar system}
G. J. Olmo, Phys. Rev. \textbf{D}, 72, 083505 (2005);
V. Faraoni, Phys. Rev. \textbf{D}, 74, 023529 (2006).

\bibitem{Amendola07}
L. Amendola, R. Gannouji, D. Polarski, S. Tsujikawa, Phys. Rev. D, \textbf{75}, 083504 (2007).

\bibitem{cosm perturb}
S. M. Carroll, I. Sawicki, A. Silvestri, M. Trodden, New J. Phys. \textbf{8}, 323 (2006);
Y. S. Song, W. Hu, I. Sawicki, Phys. Rev. D, \textbf{75}, 044004 (2007);
R. Bean, D. Bernat, L. Pogosian, A. Silvestri, M. Trodden, Phy. Rev. D,  \textbf{75}, 064020 (2007).


\bibitem{lcdm limit}
T. Chiba, T. L. Smith, A. L. Erickcek, Phys. Rev. D, \textbf{75}, 124014 (2007);
J. Dossett, B. Hu, D. Parkinson, J. Cosm. Astrop. Phys., \textbf{03}, 046 (2014);
W. Hu, I. Sawicki, Phys. Rev. D, \textbf{76}, 064004 (2007);
S. A. Appleby, R. A. Battye, Phys. Lett. B, \textbf{654}, 7 (2007).

\bibitem{Martins17}
C. J. A. P. Martins, Rep. Prog. Phys., \textbf{80}, 12 (2017).

\bibitem{Dunsby16}
P. K. S. Dunsby, O. Luongo, Int. J. Geom. Meth. Mod. Phys., \textbf{13}, 1630002 (2016).




\end{thebibliography}
\end{document}